\documentclass[12pt,a4paper]{article}
\usepackage{amsmath}
\input epsf
\usepackage[dvips]{graphicx}
\usepackage{times}

\begin{document}
\begin{titlepage}
\title{On the large-$t$ elastic scattering at $\sqrt{s}=7$ TeV}
\author{ S.M. Troshin,
  N.E. Tyurin\\[1ex] \small\it Institute
\small\it for High Energy Physics,\\\small\it Protvino, Moscow Region, 142281 Russia}
\date{}
\maketitle
\begin{abstract}
We discuss discrimination of the scattering mechanisms on the basis of 
functional dependencies of the large-$t$ elastic scattering  suggested by  the recent
data from the TOTEM experiment. It is shown that Orear exponent is  in a better agreement with the 
data than the power-like dependence used by the TOTEM. This implies that the collective dynamics is dominating  
over   the point-like mechanism related to the scattering of the proton constituents. 

We also emphasize that vanishing of the helicity-flip
amplitudes contributions at the LHC energies would result in  appearance of the dip--bump structures at higher values of transferred
momenta.  

\end{abstract}
\vfill
\end{titlepage}

\section*{Introduction}
The experiments at the LHC have already delivered many interesting results related to the processes of the elastic scattering  
and multiparticle production (cf. e.g.  \cite{totem,lhc, lhc1,lhc2}).  
Presence of the significant collective effects has been revealed  in the inelastic $pp$-interactions. 
Those effects can be interpreted as a manifestation 
of the high degree  coherence in the $pp$-collisions resulting in the multiparticle final states \cite{intjrev}. 

On the other hand, the initial particles preserve their identity after the interaction
 in elastic scattering and in this sense the studies of the elastic scattering 
provide a complimentary relevant instrument. It can also lead to a new
insight on the nature of the nonperturbative hadronic interactions, role of confinement 
 and  possible asymptotic mechanism of hadron scattering. 

The studies of the correlations of the large-$t$ elastic scattering  alongside with the multiparticle production studies
can bring a useful information on the hadron scattering at very high energies \cite{plb}. These correlations can serve as an independent and
additional qualitative experimental signatures of the possible asymptotic mechanism.

The highest energy where differential cross-section of elastic scattering has been measured was $\sqrt{s}=7$ TeV \cite{totem}.
These measurements have demonstrated existence of the dip and smooth decreasing dependence on the transferred momentum in the region
beyond the second maximum. Similar smooth dependence with lower value of  slope has been observed earlier in the 
wide $t$-region at the CERN ISR and at lower energies as well \cite{isr,hart}.

Low-$t$ region and dip structure are commonly associated  with the  diffraction on the opaque object.  Thus, in
this region the main role is to be attributed to the nonperturbative collective dynamics.  

In the region beyond the second maximum,  additional
dips and bumps are absent.  Such dependence is interpreted in different  ways. It can be considered as a result of the composite  hadron structure.
In this case a power-like dependence \cite{qcr,qcr1,dl} is to be used as a relevant function for the description of the experimental data. 
Contrary,  the exponential  form can also be applied. The above--mentioned  dependencies imply different dynamical mechanisms, namely, 
power--like behavior corresponds to the composite scattering dynamics where coherence is absent and point--like 
constituents are independent, while the exponential form can be
associated with coherent collective interactions  persisting at large values of $-t$.

In this Letter we discuss the issue of the coherence in the large-$t$ elastic scattering in the light of the new experimental data at $\sqrt{s}=7$ TeV and
demonstrate that the available data are in better agreement with predictions based on the collective dynamics.

\section{Functional dependence of $d\sigma /dt$ at large $t$}
As it was mentioned in the Introduction, the differential cross--section of elastic scattering has been
measured at various values of $-t$ at the LHC energy $\sqrt{s}=7$ TeV.  
The interest in the large-$t$ region is related to the possible detection
of the point--like dynamics in the proton scattering \cite{dl}. The dependence $ d\sigma /dt \sim |t|^{-8}$ results from the perturbative 
QCD 3-gluon-exchange
diagrams (cf. \cite{dl}). From the theoretical point of view, this dependence and its apparent agreement with the experimental data
imply dominance of the  pinch singularities of the Landshoff type independent--scattering diagrams   despite a presence
of the Sudakov form factor  \cite{pire} in the 
elastic scattering in this region of the transferred momenta.

The power-like parametrization 
\begin{equation}\label{dl}
d\sigma/dt \sim |t|^{-7.8}
\end{equation}
was used for  description of the experimental data in the region between 1.5 $(GeV)^2$ and 2.0  $(GeV)^2$ in the experimental paper \cite{totem} and 
it corresponds to the $|t|^{-8}$ dependence with account of the experimental errors. 
This dependence is depicted on the Fig. 1 (dashed line).
\begin{figure}[hbt]\label{78}
\begin{center}
\resizebox{10cm}{!}{\includegraphics*{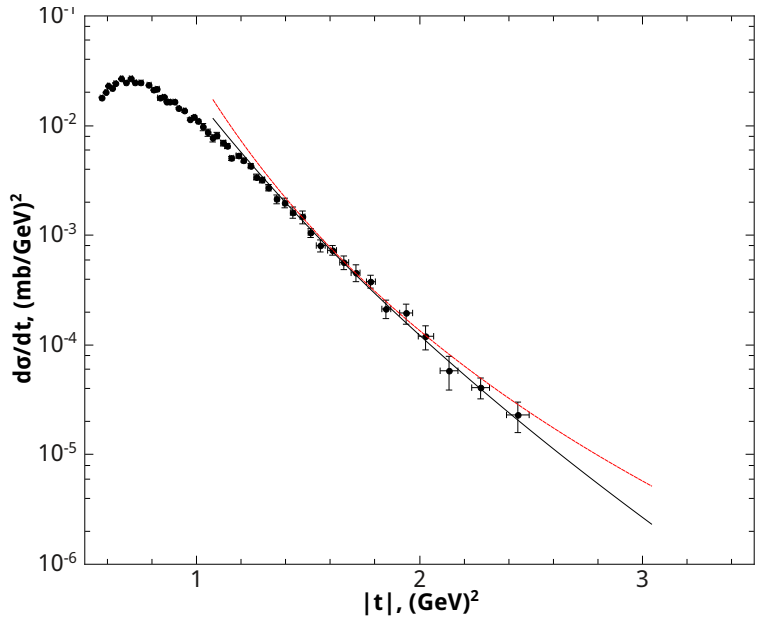}}
\end{center}
\vspace{-1cm}
\caption{Dependence of the large-$t$ elastic scattering differential cross-section.}
\end{figure}
At the LHC energy $\sqrt{s}=7$ TeV Eq. (\ref{78}) allows to fit data in the rather narrow region of the transferred momenta.

Note that the independent--scattering diagrams are not rotationally--invariant and  therefore presence of the orbital momentum
can lead to the specific spin effects in elastic scattering which can be used as a signature of the presence or absence 
of such contribution \cite{pira}. Vanishing of the helicity non-conserving dynamics with energy  disagrees with the survival  of the above
mechanism. 

 Another dependence, of an exponential type, called Orear behavior \cite{orear} 
\begin{equation}\label{or}
d\sigma/dt \sim\exp(-c_{\mbox{o}}\sqrt{-t}) 
\end{equation}
can also describe  the experimental data  with $c_{\mbox{o}}\simeq 12$  $(GeV)^{-1}$, cf. Fig. 1 (solid line). 
This value of the slope parameter  is about twice as much larger compared to the corresponding value of  
$c_{\mbox{o}}$ at the CERN ISR and at lower energies \cite{hart}. It will be shown further that the energy dependence of the slope can result from 
the presence of phase in $U$--matrx or double helicity-flip amplitudes contribution.

It is evident that the exponential dependence on $\sqrt{-t}$ 
 describes  experimental data in the wider region of $-t$-values and use of the power-like dependence
for the data analysis seems to  be premature and  misleading.
The  measurements at $-t\geq 2.5$ $(GeV)^{2}$ would allow to clarify the experimental situation.

At the moment however, one can conclude that the coherent scattering is a dominating one in the region of the large $-t$-values.
It should be noted that the experimental data for elastic large-angle scattering at lower energies were analysed in \cite{hendry}
 and conclusion on the presence of substantial 
coherent--scattering effects even at large angles has been done.

\section{Coherent dynamics in the large-$t$ elastic scattering}
Several dynamical mechanisms lead to the Orear dependence but all of them have collective non-perturbative dynamics at the origin. 
 Such behavior can result from the
 presence of the poles of the scattering amplitude in the complex plane of the impact parameter. These moving Regge-like poles
in the direct channel of reaction appear in the tunneling picture of hadronic diffraction where they are related to the surface creep waves
\cite{schremp} and match the concept of the geometrical picture where resonances are peripherally excited at the edge of the interaction
 region \cite{schremp1}.

The appearance of such poles can also be a result \cite{ech} 
 of the rational form of the scattering amplitude unitarization. Such form of the amplitude unitarization might be related with the confinement
phenomena \cite{mplc}. In this form of unitarization the elastic scattering $S$--matrix element ($2\to 2$)  can be written in the form
\begin{equation}\label{um}
 S(s,b)=\frac{1-U(s,b)}{1+U(s,b)}
\end{equation}
in the case of pure imaginary $U$--matrix ($U\to iU$). 
Elastic scattering amplitude has the poles in the complex impact parameter plane whose position is determined by the solutions of the equation
\begin{equation}\label{pol}
1+U(s,b)=0
\end{equation}
Solutions of this equation depend weakly on the particular form of the function $U$ (provided it has known analytical properties in the complex 
$t$-plane)\footnote{This solution is valid in the case when $U(s,b=0)>1$.}:
\begin{equation}\label{polp}
 b_m(s)=R(s)+i\frac{\pi}{\mu}m,
\end{equation}
where $m$ is the positive integer number and $R(s)$ is the hadron interaction radius.  In the case when the function $U$ has the phase 
$\alpha(s)\neq \pi/2$ 
the elastic scattering amplitude $F(s,t)$ can be expanded over parameter $\tau (t)=\exp(-\pi\sqrt{-t}/\mu)$ in the following form:
\begin{equation}\label{exp}
F(s,t)\sim s\sum_{m=1}^\infty \tau^m(t)\varphi_m[R(s), \alpha(s), \sqrt{-t}],
 \end{equation}
where $\varphi_m$ are the smooth functions of $\sqrt{-t}$ if the function $\alpha(s)$ is not identically equal to $\pi/2$ and oscillating  ones
in the case $\alpha(s)= \pi/2$ (details can be found in \cite{ech,ph}).  At large values of $-t$ in the expansion Eq. (\ref{exp}) one can keep only several
or even one term and arrive to the Orear dependence Eq. (\ref{or})\footnote{ Parameter $\mu$ in the expression for the expansion parameter 
$\tau$ can be related to the masses of the constituent quarks\cite{ph} and parameter  $c_{\mbox{o}}$ in Eq.(\ref{or})
 can aquire a nontrival energy dependence due to the phase   $\alpha(s)$.}. 

\section{Role of spin in the large-t differential cross--section oscillations}
Thus, in the case of pure imaginary scattering amplitude the poles in the impact parameter plane provide additional oscillating factors in front
of the exponential dependence in the amplitude.  These oscillations are common for the diffraction picture with pure imaginary scattering amplitude.
It is reasonable to expect that the scattering amplitude at very high energies will tend to the pure imaginary function. 
Therefore the absence of the oscillations in the region of large $-t$   which can naturally be explained at lower energies
 by the significant role of the generalized reaction matrix phase  would not work
at the LHC energies. 

Another explanation of the smooth dependence of the differential cross--section is  the role
of the spin effects associated with the presence of the essential double helicity-flip amplitude contribution \cite{dbf}. 
It should be noted that the helicity--flip amplitudes can also be represented in the form of the expansion similar to Eq. (\ref{exp})
These helicity amplitudes (there are five independent helicity amplitudes for $pp$-scattering) can be written in the form
\begin{equation}\label{hel}
 F_i(s,t)={\phi}_i(s,t)\cos [R_i(s)\sqrt{-t}+\omega_i(s)],
\end{equation}
where the functions ${\phi}_i(s,t)$ are exponentially decreasing with $\sqrt{-t}$.
It has been shown that  the double helicity--flip amplitudes $F_2$ and  $F_4$ are important at large values of $-t$
and compensates oscillations of the helicity non-flip amplitudes \cite{dbf}. 

However, again it is difficult to expect that any helicity-flip 
amplitudes would survive at such high energies as the LHC energies are.   Vanishing  of the helicity-flip amplitudes 
 would, therefore, result in appearance
of the  oscillations at higher $-t$-values provided, of course, that the above spin mechanism is mainly 
responsible for the smooth large-$t$ dependence at lower energies.
Thus, a possible appearance of the above oscillations in the differential cross-section at higher values of $-t$
 should be interpreted then as an observation of the $s$-channel helicity conservation 
in $pp$-scattering at the LHC energies.

\section*{Conclusion}
Thus, we would like to emphasize that use of the power--like dependence for the description of the experimental data for the large--$t$
elastic scattering at the LHC energy $\sqrt{s}=7$ $TeV$ seems to be premature in the currently available for the experimental measurements
region of the transferred momenta. There are theoretical and experimental arguments in favor of the expectation that the Orear dependence 
(with or without oscillating factors) being more relevant for the descriprion of the elastic scattering in this kinematical region of  large
 energy and transferred momentum values.

We would like to note  that the
presence of the possible oscillations of the differential cross-section at higher values of $-t$ 
cannot be excluded on the current experimental basis as it was mentioned in the 
recent paper \cite{uzh}. The papers  \cite{totem,uzh}
provide also  a list of  references to the several quantitative models of elastic scattering not mentioned  in the present note. 
We concentrated here on the  qualitative dynamical implications of large-$t$ experimental data for $pp$-elastic 
measured at the LHC energy $\sqrt{s}=7$ TeV.

\end{document}